\documentclass[smallextended]{svjour3my}
\smartqed
\usepackage{graphicx}

\usepackage{hyperref}
\journalname{General Relativity and Gravitation}
\begin{document}


\title{Regular frames and particle's rotation near a black hole}

\titlerunning{Regular frames and particle's rotation near a black hole}

\author{Yuri V. Pavlov$^{1,2}$        \and
        Oleg B. Zaslavskii$^{2,3}$}

\authorrunning{Yu.V. Pavlov \and O.B. Zaslavskii}

\institute{Yu. V. Pavlov \at
              \email{yuri.pavlov@mail.ru}
           \and
           O. B. Zaslavskii \at
              \email{zaslav@ukr.net}
           \and
${}^{1}$\,Institute of Problems in Mechanical Engineering,
Russian Academy of Sciences,\\
61~Bol'shoy pr., St. Petersburg 199178, Russia
           \\ \\
${}^{2}$\,N.I.\,Lobachevsky Institute of Mathematics and Mechanics,
    Kazan Federal University, 18 Kremlyovskaya St., Kazan 420008, Russia;
           \\ \\
${}^{3}$\,Department of Physics and Technology, Kharkov V.N.\,Karazin
National University, 4~Svoboda Square, Kharkov 61022, Ukraine}

\date{Received: 28 September 2018 / Accepted: 29 April 2019}

\maketitle

\begin{abstract}
We consider a particle moving towards a rotating black hole.
We are interested in the number of its revolution $n$ around a black hole.
In our previous work (Pavlov and Zaslavskii in
\href{https://doi.org/10.1007/s10714-017-2333-5}
{Gen Relativ Gravit 50: 14, 2018}.
\href{https://arxiv.org/abs/1707.02860}{arXiv:1707.02860})
we considered this issue in the Boyer-Lindquist type
of coordinates with a subsequent procedure of subtraction.
Now, we reconsider this issue using from the very beginning the frames
regular on the horizon.
For a nonextremal black hole, regularity of a coordinate frame leads to
the finiteness of a number of revolutions around a black hole without
a subtraction procedure.
Meanwhile, for extremal black holes comparison of $n$ calculated in the regular
frame with some subtraction procedures used by us earlier shows that
the results can be different.


\end{abstract}

\section{\label{sec1}Introduction}

It is well known that if a particle falls towards a black hole, there is a
sharp contrast between the time intervals measured by an observer at
infinity ($t$) and an observer comoving with the particle (the proper time $%
\tau $). To reach the event horizon, a particle needs an infinite~$t$.
Meanwhile, $\tau $ (except from some special situations) is finite. The
similar relation exists between the number of revolutions~$n$ around a
rotating black hole~\cite{rev}. For a remote observer, it is always
infinite. For a comoving observer it can be finite or infinite.

In the previous work~\cite{rev} we showed that the relationship between
these two characteristics can be quite nontrivial. In particular, it can
depend not only on the type of black hole and a trajectory but also on a way
the measurement are performed. We considered (i) the measurement for a
free-falling particle with respect to another free-falling one, (ii)
measurements with respect to a particle that is not in free fall, (iii) with
respect to a black hole itself. In all three cases it was impossible to view
the effect using one fixed coordinate frame.

The divergences of $n$ are connected with the fact that the original
coordinates (like the Boyer-Lindquist ones for the Kerr metric) fail near
the horizon, so both $t$ and the polar angle variable $\phi $ diverge.
Meanwhile, in recent years, coordinate frames regular near the rotating
black holes were constructed \cite{dor}--\cite{gen} that enabled one to
trace rather subtle details of particle behavior near the horizon~\cite%
{ted11}, \cite{near} or give physical interpretation of metrics in terms of
rotating fluid~\cite{fluid}.  The aim of the present work is to consider the
properties of revolution of a particle around a black hole using this type
of angle variable. Then, there is no need to develop special schemes for
subtraction of the contribution to $n$ from the reference frame since this
subtraction is already implicitly contained in the definition of a ``good''
angle variable. We will see that this simple and direct procedure agrees
with the aforementioned method (i) that, however, can be different from (ii)
and (iii).

It turned out that overlap between different but related issues (description
of geometry near the horizon and description of particle revolution around
it) lead us to construction of the rotational analogue of the
contracting/expanding Eddington-Finkelstein (or Lema\^{\i}tre) frames for
black/white holes.  This generalizes corresponding construction, known for
the Kerr or Kerr-Newman metric.

\section{\label{sec2}Metric and equations of motion}

Let us consider the metric
\begin{equation}
ds^{2}=-N^{2}dt^{2}+g_{\phi }(d\phi -\omega dt)^{2}+\frac{dr^{2}}{A}
+g_{\theta }d\theta ^{2} .  \label{met}
\end{equation}
Here, the event horizon is described by $N=0$. We assume that the metric
coefficients do not depend on $t$ and $\phi $, so that the energy $E$ and
angular momentum $L$ of a particle are conserved. We also assume
\begin{equation}
N^{2}=\alpha \Delta , \ \ A=\frac{\Delta }{\rho ^{2}} ,  \label{abn}
\end{equation}
where $\Delta =\Delta (r)$, $\alpha >0$ and $\rho >0$ depend on both~$r$ and~%
$\theta $.  The maximum zero $r=r_{+}$ of $\Delta $ corresponds to the event
horizon, $\alpha $ and $\rho $ are supposed to remain finite and nonzero at $%
r=r_{+}$.

In what follows, we use notation
\begin{equation}
X=E-\omega L.  \label{x}
\end{equation}
It can be written in a more general form as $X=-mu_{\mu}(\kappa^{\mu} +
\omega\zeta^{\mu})$, where $\kappa^{\mu}$ is the Killing vector responsible
for time translations and $\zeta^{\mu}$ is that responsible for rotations.
We restrict ourselves by motion in the equatorial plane $\theta =\pi /2$
only.

Equations of geodesic motion read
\begin{equation}
m\frac{dt}{d\tau }=\frac{X}{N^{2}} ,
\end{equation}
\begin{equation}
m\frac{d\phi }{d\tau }=\frac{L}{g_{\phi }}+\frac{\omega X}{N^{2}} ,
\label{angle}
\end{equation}
\begin{equation}
m\rho \sqrt{\alpha}\,\frac{dr}{d\tau }= -Z , \ \ Z=\sqrt{X^{2}-N^{2} \left(
\frac{L^{2}}{g_{\phi }}+m^{2} \right)},  \label{ro}
\end{equation}
where we assumed that a particle moves towards a black hole, so $dr/d\tau <0$%
.  Here, $m$ is a particle's mass, $\tau $ being a proper time.  It follows
from~(\ref{angle}), (\ref{ro}) that
\begin{equation}
\frac{d\phi }{dr}=-\frac{\rho \sqrt{\alpha }}{Z} \left( \frac{L}{g_{\phi }}+%
\frac{\omega X}{N^{2}}\right) ,  \label{phr}
\end{equation}
\begin{equation}
t=-\int^{r}\! \frac{X\rho\, dr^{\prime}}{\sqrt{\alpha }\Delta Z} .
\label{tr}
\end{equation}

We imply that the so-called forward-in-time condition $dt/d\tau >0 $ is
satisfied because of which $X\geq 0$.

\section{\label{sec3}Coordinate transformation to regular frames}

In~\cite{gen}, a general approach was suggested that enables one to build
frames regular near the horizon.  This includes previous known coordinate
frames, in particular, Painlev\'{e}-Gullstrand ones for the Kerr and
Kerr-Newman metrics~\cite{dor}--\cite{lin}.  To make presentation
self-contained, we repeat below some general formulas from~\cite{gen}, where
a reader can find further details.  For the metric~(\ref{met}),
transformations within the equatorial plane have the form
\begin{equation}
dt=d\bar{t}+\frac{z(r)dr}{\Delta } ,  \label{t}
\end{equation}
\begin{equation}
d\phi = d\bar{\phi}+\frac{\xi(r,\theta)\, dr}{\Delta }+ \delta(r,\theta)\,
d\theta .  \label{phi}
\end{equation}
The key property of barred coordinates consist in that (apart from some
exceptional cases) the trajectories of particles falling on a black hole
reach the horizon for a finite time $\bar{t}$ and perform a finite number of
revolutions. This is discussed below in Sec.~\ref{sec6}.

    We restrict ourselves by motion of particles in the equatorial plane
$\theta =\mathrm{const}=\pi /2$, so the last term in~(\ref{phi})
is irrelevant and put $\delta =0$.
It follows from (\ref{t}), (\ref{phi}) that
\begin{equation}
d\phi -\omega dt=d\bar{\phi}-\omega d\bar{t}+h dr ,
\end{equation}
\begin{equation}
g_{rr}=\mu +g_{\phi} h^{2} ,
\end{equation}
    where we introduced new functions $h$ and $\mu $ according to
\begin{equation}
h\Delta =\xi -\omega z ,  \label{2}
\end{equation}
\begin{equation}
\mu \Delta =\rho^{2}-z^{2}\alpha .  \label{ksi}
\end{equation}
In new variables, the metric has the form
\begin{equation}
ds^{2}=-N^{2}d\bar{t}^{2}-2z\alpha drd\bar{t}+g_{\phi }(d\bar{\phi}-
\omega d \bar{t}+hdr+d\theta \delta )^{2}+\mu dr^{2}.
\label{metric2}
\end{equation}

We want to kill the divergences in the metric coefficient $g_{rr}$.
To this end, we choose $h$ and $\mu $ finite on the horizon.
If we specify some functions $z(r)$ and $h(r,\theta )$ where $z(r)$ is
also finite on the horizon, it follows from~(\ref{2}), (\ref{ksi}) and
the fact that $\Delta =0$ on the horizon, that
\begin{equation}
(\xi -\omega z)_{r=r_{+}}=0,  \label{omhor}
\end{equation}
\begin{equation}
\left( z^{2}\alpha -\rho ^{2}\right) _{r=r_{+}}=0.  \label{za}
\end{equation}

Then,
\begin{equation}
\frac{d\bar{\phi}}{dr} = -\frac{\xi }{\Delta }-\frac{\rho \sqrt{\alpha }}{Z}
\left( \frac{L}{g_{\phi }}+\frac{\omega X}{N^{2}} \right) = -\frac{\omega }{
\Delta } \left( z+ \frac{ X\rho }{Z\sqrt{\alpha }} \right) -h -\frac{\rho
\sqrt{\alpha }}{Z}\frac{L}{g_{\phi }}.  \label{phi1}
\end{equation}
The number of revolutions that particle experiences during travel between
points 1 and 2 is equal to
\begin{equation}
n = \frac{\Delta \bar{\phi}}{2\pi } , \ \ \Delta \bar{\phi}=\bar{\phi}_{2}-
\bar{\phi}_{1} .  \label{n}
\end{equation}

It is worth paying attention to the choice
\begin{equation}
z=-\frac{\rho \sqrt{1-\alpha \Delta }}{\alpha },  \label{zd}
\end{equation}%
\begin{equation}
h=\frac{\omega \rho ^{2}}{z},\ \ \mu =\alpha \rho ^{2},
\end{equation}%
\begin{equation}
\xi =\frac{\omega \rho ^{2}}{z\alpha }.  \label{ksik}
\end{equation}%
Then, for a particle moving with $L=0$, $E=m$ we see the angle
$\bar{\phi}=\mathrm{const}$.
This is the generalization of the corresponding property~\cite{ted11}
inherent to the Kerr metric in coordinates of Ref.~\cite{dor}.
One reservation is in order. The aforementioned choice implies that the
combination~(\ref{zd}) does not contain the angle~$\theta $. This is valid
for the Kerr metric but not necessarily in a general case. However, for
motion in the plane $\theta =\pi /2=\mathrm{const}$ the ansatz~(\ref{zd}) is
sufficient.

\section{\label{sec4new}General criteria}

We want to elucidate, whether the change of the angle variable
$\Delta \bar{\phi} $ during a particle fall into a black hole
is infinite or finite.
To this end, we examine the behavior of the time variable $\bar{t} $ and show
that in cases under consideration it remains finite.  Then, we calculate the
horizon limit of the quantity $\frac{d\bar{\phi}}{d\bar{t}} $.
As $\bar{t} $ changes in finite limits, a finite
$\left( \frac{d\bar{\phi}}{d \bar{t}} \right)_{H} $ is quite sufficient
to conclude that $\Delta \bar{\phi} $ is
finite as well (hereafter, subscript ``$H$'' means that a corresponding
quantity is calculated on the horizon).  Alternatively, one may evaluate the
quantity $\left( \frac{d\bar{\phi}}{d r}\right)_{H} $ since $r $ is supposed
to change within a finite interval between some initial value $r_{1}$ and
the horizon radius $r_{+}$.

Let us consider now different types of black holes and of particles
separately.

\section{\label{sec4}Nonextremal black hole}

Now, near the horizon the expansion of $\omega $ takes the form~\cite{dirty}
\begin{equation}
\omega =\omega_{H}- B_{1}N^{2}+ O(N^{3}).
\end{equation}
For the Kerr-Newman metric, $B_{1}>0$.

Then, it follows from (\ref{x}) and (\ref{ro}) that near the horizon
\begin{equation}
X=X_{H}+B_{1}N^{2}L+O(N^{3}) ,  \label{xh}
\end{equation}
\begin{equation}
Z\approx X-\frac{N^{2}}{2X_{H}} \left( \frac{L^{2}}{g_{H}}+m^{2} \right) .
\label{zh}
\end{equation}
We have from (\ref{za}) that near the horizon either $z+ \rho / \sqrt{\alpha
} \approx 0$ or $z - \rho / \sqrt{\alpha} \approx 0$.  Choosing the first
option, we have near the horizon
\begin{equation}
z+\frac{\rho }{\sqrt{\alpha }}\approx -b\Delta ,  \label{zb}
\end{equation}
where $g_{H}=g_{\phi }$ $(N=0)$ and $b$ is the model-dependent coefficient.
Expansion~(\ref{zb}) agrees with~(\ref{za}).  Then,
\begin{eqnarray}
\left( \frac{d\bar{\phi}}{dr}\right)_{H} &=& \omega_{H}b-h_{H}
- \frac{\rho_{H}\sqrt{\alpha_{H}}}{X_{H}} \frac{L}{g_H}
- \omega_{H} \frac{\rho_{H}\sqrt{\alpha_{H}}}{2 X_{H}^2}
\left( \frac{L^{2}}{g_{H}} + m^{2} \right).
\label{d}
\end{eqnarray}
It is finite. It follows from the above equations that the quantity
$\bar{\phi} $ obtained by the integration of the right hand side
of~(\ref{phi1}) is also finite.
Thus the angle changes at a finite value during infall of a particle
to a black hole.

\section{\label{sec5}Extremal black hole}

Now
\begin{equation}
N^{2}=D(r-r_{+})^{2}+O((r-r_{+})^{3}) ,
\end{equation}
where $D>0$ is some constant. The expansion for $\omega $ takes the form
\cite{dirty}
\begin{equation}
\omega =\omega_{H}-B_{1}N+O(N^{2}) .  \label{ome}
\end{equation}

Below, we use classification of particles according to which a particle with
$X_{H}>0$ is called usual and that with $X_{H}=0$ is called critical.
Here, $X$ is used according to definition~(\ref{x}).

\subsection{\label{sec51}Usual particles}

    Then, one can see that the expression for~(\ref{d}) on the horizon retains
its validity, so $\bar{\phi}$ is finite.

\subsection{\label{sec52}Critical particles}

Using~(\ref{ome}), one obtains
\begin{equation}
X=B_{1}LN+O(N^{2}) ,
\end{equation}
\begin{equation}
Z = Z_{1}N+O(N^{2}), \ \ Z_{1} = \sqrt{\frac{E^{2}}{\omega_{H^{2}}} \left(
B_{1}^{2}-\frac{1}{g_{H}} \right) - m^{2}} .
\end{equation}
As a result,
\begin{equation}
n\approx \frac{\rho_{H}^{2} \sqrt{ \alpha_{H}} \omega_{H} \left(
\frac{B_{1}E}{ \omega_{H}} -Z_{1} \right)}{2\pi Z_{1}D(r-r_{+})} ,
\label{c}
\end{equation}
so~(\ref{c}) diverges. Equation~(\ref{c}) corresponds to Eq.~(48) of~\cite{rev}.

\section{\label{sec6}Two types of frames: rotational analogues of expanding and
contracting frames}

We saw that, after introducing a regular frame, the number of revolutions
becomes a finite, except a rather special case of the critical particle
moving around the extremal horizon. Meanwhile, one can notice here a rather
interesting peculiarity. Near the horizon, Eq.~(\ref{za}) admits two
branches for~$z$.  Regularization of $n$ leads to a finite result for a
quite definite choice of a sign of $z$ near the horizon. However, if instead
of~(\ref{zb}) we take $z\approx + \rho / \sqrt{\alpha }$, the quantity~$n$
remains infinite, although the metric itself looks regular since~(\ref{za})
is satisfied.  How can it happen?

To elucidate this issue, let us consider Eq.~(\ref{t}). It follows from it
that
\begin{equation}
\bar{t}=t-\int_{r_{1}}^{r}\frac{z(r^{\prime })dr^{\prime }}{\Delta
(r^{\prime })},  \label{ttz}
\end{equation}%
where $r_{1}$ is some constant. Let a free particle fall towards a black
hole. Using equation of motion~(\ref{tr}), we have, choosing the integration
constant properly,
\begin{equation}
\bar{t}=-\int_{r_{1}}^{r}\frac{dr^{\prime }}{\Delta }\left( z+\frac{X\rho }{Z%
\sqrt{\alpha }}\right) ,  \label{tz}
\end{equation}%
where $\bar{t}(r_{1})=0$. A particle moves with decreasing $r$, so further $%
r<r_{1}$, $t>0$.

For definiteness, let us consider a nonextremal black hole.  Then, near the
horizon, $X / Z \approx 1$ as is seen from~(\ref{zh}).  Correspondingly, it
is the choice~(\ref{zb}) that enables us to obtain a finite value for $\bar{t%
}$.  In other words, a particle moving from the outside towards a horizon,
reaches it for a finite interval of~$\bar{t}$, making a finite number of
revolutions.

In a similar way, we can consider a particle that moves outward.  Then, $dr
/ dt > 0 $ and, instead of~(\ref{tz}) we have
\begin{equation}
\bar{t} = - \int_{r_{1}}^{r}\frac{dr^{\prime }}{\Delta } \left( z-\frac{%
X\rho }{Z\sqrt{ \alpha }} \right).
\end{equation}
With the same choice~(\ref{zb}) we obtain divergent $\bar{t}$.  Also, the
angle variable diverges if $r\rightarrow r_{+}$.

This has a clear analogy with the contracting and expanding
Eddington-Finkelstein or Lema\^{\i}tre frames (see, e.g. Sec. 33 in~\cite%
{mtw}, Sec. 2.4 and 2.5 of~\cite{fn}).  The contracting frame describes
properly a history of particles falling under the horizon (black hole) but
is unable to describe the history of particle appearing from the inner
region (white hole). For the expanding system, the situation is opposite.
The aforementioned frames are suited for the description of radial motion.
Meanwhile, now we dealt with a similar problem for a rotational motion.

The finiteness of $\bar{t}$ gives one more simple explanation of finiteness
of the number of revolutions.  It follows from Eq.~(\ref{d}) that $\frac{d%
\bar{\phi}}{dr}$ is finite.  As $\frac{dr}{d\bar{t}}$ is finite as well, $%
\frac{d\bar{\phi}}{d\bar{t}}$ is finite too.  Then, the barred angle changes
at a finite value during particle travel to the horizon.

\section{\label{sec7}Dragging effect for particle scattered by a black hole}

    We want to stress the following interesting property of a particle that is
scattered by a black hole. Let such a particle move from large $r_{1}$ to
some minimum distance $r_{f}$ and, afterwards, escapes with the same
parameters $m, E, L$ to infinity again.
Then, the change of the angle from $r_{1}$ to $r_{f}$ is equal to
\begin{equation}
\Delta _{1}\phi =+\int_{r_{f}}^{r_{1}}\eta (r^{\prime })\,dr^{\prime },
\label{df}
\end{equation}
where $\eta =\frac{\rho \sqrt{\alpha }}{Z}\left( \frac{L}{g_{\phi }} +
\frac{\omega X}{N^{2}}\right) $ according to~(\ref{phr}).
    From $r_{f}$ to $r_{1}$ a particle moves with $dr/dt>0$,
so that the sign in Eq.~(\ref{phr}) changes.
    Therefore, if  a particle returns to the same $r_{1}$, it receives
an additional change expressed by the same formula~(\ref{df}).
As a result, the full change $\Delta \phi =2\Delta _{1}\phi $.
If $r_{f}\rightarrow r_{+}$, $\Delta \phi \rightarrow \infty $ due to
the factor $N^{2}$ in the denominator of the second term in~$\eta $.

If, instead of $\phi $, one uses $\bar{\phi}$, one can easily see
from(\ref{phi}) that the change of the angle $ \bar{\phi} $
for moving from $r_1$ to $r_f$ and return is
\begin{equation}
\Delta \bar{\phi} =\int_{r_{f}}^{r_{1}} \left( \frac{\xi }{\Delta } +
\eta (r^{\prime })\right) dr^{\prime }
+ \int_{r_{1}}^{r_{f}} \left( \frac{\xi }{\Delta } - \eta (r^{\prime })
\right) dr^{\prime} .
\label{dfdop}
\end{equation}
    In~(\ref{dfdop}) two integrals containing $\xi $ mutually cancel,
and again we obtain $\Delta \bar{\phi}=2\Delta_{1}\phi $,
where $\Delta_{1}\phi $ is given by Eq.~(\ref{df}).
In this sense, the considered coordinate transformation does
not change this result.
What is especially interesting is that the dragging effect persists and is
described by the same formula $\Delta \bar{\phi} = 2\Delta _{1}\phi $ even if,
by choosing appropriate functions $z(r)$, $h(r),
$ we can achieve $\bar{\phi}=\mathrm{const}$ for the falling in equatorial
plane. (But, in the chosen coordinates, the angle variable will not remain
constant during motion in the outward direction.)

\section{\label{sec8}Kerr metric}

    Let us consider the Kerr metric as an explicit example.
    Then, in the Boyer-Lindquist coordinates
    \begin{equation}
\Delta =r^{2}-2Mr+a^{2}, \ \ \alpha =\frac{1}{\Sigma } , \ \
g_{\phi }=\Sigma, \ \ g_{\theta} =\rho ^{2},
\end{equation}
    \begin{equation}
\rho ^{2}=r^{2}+a^{2}\cos ^{2}\theta , \ \
\Sigma =r^{2}+a^{2}+\frac{ 2Mra^{2}}{\rho ^{2}}\sin ^{2}\theta ,
\end{equation}
    \begin{equation}
\omega =\frac{2Mar}{\rho ^{2}\Sigma}, \ \
\rho^{2}\Sigma =(r^{2} + a^{2})^{2} - \Delta a^{2}\sin ^{2}\theta ,
\end{equation}
    where $M$ is a black hole mass, $aM$ being its angular momentum.
    The event horizon lies at $r = r_+ = M + \sqrt{M^2-a^2}$.
    The black hole angular velocity
    \begin{equation}
\omega_{H}=\left.\omega\right|_{\Delta=0}=\frac{a}{r_{+}^{2}+a^{2}} .
\end{equation}

    The choice
    \begin{equation}
z=-\sqrt{2Mr(r^{2}+a^{2})},
\end{equation}
    \begin{equation}
h=\frac{\omega \rho ^{2}}{z} =
- \frac{a\sqrt{2Mr}}{\Sigma \sqrt{r^{2}+a^{2}}},
\end{equation}
    \begin{equation}
\xi = -\frac{\sqrt{2Mr}a}{\sqrt{r^{2}+a^{2}}}
\end{equation}
    in which we took into account~(\ref{2}) corresponds to Ref.~\cite{dor}.

    The choice
    \begin{eqnarray}
h=0, \ \ z=-\sqrt{2Mr(r^{2}+a^{2})},
\nonumber \\[4pt]
\xi =-\frac{2Mar\sqrt{ 2Mr(r^{2}+a^{2})}}{\rho ^{2}\Sigma }, \ \
\mu =\frac{\rho^{2}}{\Sigma }
\end{eqnarray}
    corresponds to the Natario's frame~\cite{nat}.
    In the plane $\theta = \pi / 2$,
    \begin{equation}
\xi =-\frac{2Mar\sqrt{2Mr(r^{2}+a^{2})}}{r^{2}+a^{2}+\frac{2Ma^{2}}{r}}.
\end{equation}

    The choice
    \begin{equation}
z=-(r^{2}+a^{2}), \ \ \xi = -a
\end{equation}
    corresponds to so-called Kerr coordinates ---
see Sec. 33.2 of~\cite{mtw} or pp.~163, 164 of~\cite{he}.

\begin{figure}
\centering
\includegraphics[width=0.65\textwidth]{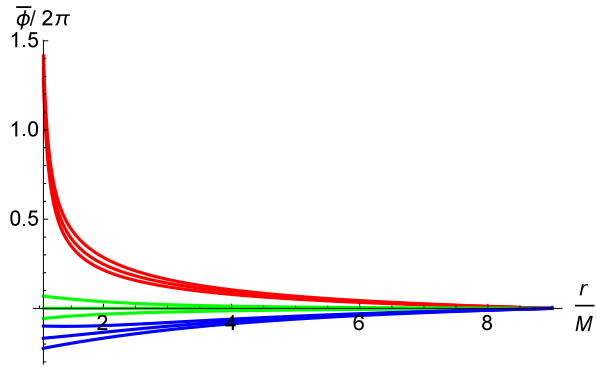}
\caption{\label{fig:epsart} The angles of rotation for particles
falling onto the black hole with $a=0.998 M$ from the point $r=9M$
with $E=m$, $L=2 mM$ (red lines), $L=0$ (green lines), and $L=-2 mM$
(blue lines) for the coordinates of Nat\'{a}rio (the top lines in groups),
Doran (the middle lines in groups), and Kerr (the lower lines in groups).}
\end{figure}

    It is also instructive to trace how changes the angle $\bar{\phi}$ during
motion of a particle when it approaches the black hole horizon.
See Fig.~\ref{fig:epsart}.

\section{\label{sec9}Discussion and conclusion}

It is instructive to compare the above results with those from~\cite{rev}.
For nonextremal black holes, the situation unambiguously agrees with Sec.~5
of~\cite{rev} and the corresponding line in Table~1 there.  For extremal
black holes, the situation is more subtle.  If a particle is usual, it is
shown in~\cite{rev} that the result depends on a way the angle is measured
(see line~3 in Table~1 there).  Now, we saw that~$n$ is finite.  For
critical particles, $n$ was found in~\cite{rev} to be infinite in all cases.
However, the asymptotic behavior is different depending on the procedure of
measurement of the angle (either as $(r-r_{+})^{-1}$ or ln($r-r_{+})$).
Now, it behaves like $(r-r_{+})^{-1}$ that agrees with Eq.~(48)
of~\cite{rev}.

Thus we see that the behavior of~$n$ (\ref{n}) in terms of $\bar{\phi}$
coincides exactly with that obtained in~\cite{rev} for measurement of
relative angles $\phi $ of two particles (this is column $n_{1}-n_{2}$ in
Table~1 there). Now $\bar{\phi}$ is a well-behaved single coordinate, so one
does not need to use the subtraction procedure. Meanwhile, now disagreement
is possible between~(\ref{n}) and other methods of computing $n$, as is seen
from the same Table~1 in Ref. \cite{rev}. In this sense, the procedure of
subtraction used in the relative measurement in~\cite{rev} is the most
natural and is confirmed now by direct calculations of the coordinate
behavior.

Also, even without referring to previous results~\cite{rev}, we can
formulate an important observation. There is an ultimate connection between
two properties: (i) the regularity of a coordinate system that can penetrate
inside across the horizon, (ii) a number of revolutions performed by a free
falling particle around the horizon. For a nonextremal black hole, a frame
can be chosen in such a way that~$n$ is always finite. It would be of
interest to extend the approach and results of the present work to the case
of nonequatorial orbits.

We want to stress that our main result concerns not the properties of
particle motion as such but has rather conceptual nature. We generalized
so-called Kerr coordinates in that our approach is applicable to generic
stationary axially symmetric black/white holes. There are two different
entities: (i) the regularity of the metric near the horizon and (ii) the
finiteness of a number of revolution during particle motion. We showed that
both properties become regular (finite) simultaneously, if a coordinate
transformation is chosen properly.

In general, the barred angles discussed in our paper involve the function of
$r$ such as $z,\xi $, $h$. In this sense, the quantity $\bar{\phi}$ is not
gauge-invariant as well as $n$. However, what is indeed gauge-invariant is
the mutual cancellation of all divergences for generic nonextremal black
holes.

For spherically symmetric black holes such as the Schwarzschild one, there
exist coordinates in which a metric is regular near the horizon, so that the
corresponding frame is able to describe trajectories of falling particles
(say, the contracting Eddington-Finkelstein or Painlev\'{e}-Gullstrand
coordinates). Meanwhile, for the description of particles appearing from a
white hole region a expanding version of such coordinates is needed. In
doing so, the new coordinates represent some combinations of previous time
and radial coordinates. In the present paper, we used the angle counterpart
of such constructions and demonstrated that the number of revolutions around
of a black hole is, as a rule, finite (some exceptional cases are also
described). We hope that, apart from the behavior of angular variables, the
corresponding frames will be useful for description of more subtle
characteristics in the spirit of the "river of space"~\cite{ham}. For
example, this will help us to build description of kinematics of particle
collision in terms of peculiar velocities that was done for radial motion
in~\cite{pec}. This is supposed to be considered elsewhere.

\begin{acknowledgements}
    This work was supported by the Russian Government Program of
Competitive Growth of Kazan Federal University.
    The work of Yu.\,P. was supported also by the Russian Foundation for
Basic Research, grant No. 18-02-00461-a.
    The work of O.\,Z. was also supported by SFFR, Ukraine, Project No. 32367.
\end{acknowledgements}


\end{document}